\documentclass[preprint,review,12pt]{elsarticle}

\usepackage{graphicx}
\usepackage{amssymb}

\usepackage{lineno}

\usepackage{color}

\journal{Journal Name}

\begin{document}

\begin{frontmatter}

\title{Effective modelling of adsorption monolayers built of complex molecules}

\author{Micha\l{} Cie\'sla}

\address{M.\ Smoluchowski Institute of Physics, Department of Statistical Physics, \\
Jagiellonian University, \L{}ojasiewicza 11, 30-348 Krak\'ow, Poland.}

\begin{abstract}
Random sequential adsorption algorithm is a popular tool for modelling structure of monolayers built in irreversible adsorption experiments. However, this algorithm becomes very inefficient when the density of molecules in a layer rises. This problem has already been solved for a very limited range of basic shapes. This study presents a solution that can be used for any molecule occupying the surface that can be modelled by any number of different disks. Additionally, the presented algorithm stops when there is no possibility to add another shape to the monolayer. This allows to study properties of fully saturated, two-dimensional random packings built of complex shapes. For instance, the presented algorithm has been used to determine the mean saturated packing fractions of monolayers built of dimers and fibrinogen. 
\end{abstract}


\end{frontmatter}


\section{Introduction}
\label{intro}

Adsorption at various interfaces has been of interest to scientists since the beginning of the last century \cite{Dabrowski2001}. This phenomenon underlies a number of important technological, environmental and biological processes. Understanding irreversible adsorption at various interfaces is of major importance for material, food and medical sciences as well as pharmaceutical and cosmetic industries. For example, protein adsorption is involved in blood coagulation, artificial organ failure, plaque formation, inflammatory response, fouling of contact lenses, ultrafiltration and membrane filtration units. It is also a prerequisite for efficient separation, purification, gel electrophoresis, and filtration of bioparticles, which is used for biosensing purposes. Therefore, modelling of monolayers built during deposition of such complex particles if of unflagging interest \cite{Adamczyk2012,Finch2013}. The most popular algorithm used in the research is random sequential adsorption (RSA) \cite{Feder1980,Evans1993}. It assumes that the motion of molecules above the interface is driven mainly by diffusion, and there is short-range attractive interaction between particle and interface and repulsive interaction between molecules themselves. The algorithm is based on consecutive iterations of the following steps:
\begin{itemize}
\item[--] a virtual molecule position and orientation is selected randomly on the interface;
\item[--] if the virtual molecule does not intersect with any other object, it is added to the interface. Otherwise, it is removed and abandoned.
\end{itemize}
The iterations are repeated many times and as a result, a set of molecules resembling adsorption monolayer is obtained \mbox{\cite{Feder1980,Onoda1986}}. The history of its application dates back to 1939 when Flory described attachment of atom groups to a vinyl polymer~\cite{Flory1939}. In 1969, Renyi found analytically the mean packing density in the car parking problem, which is one-dimensional version of RSA. Interestingly, the mean packing fraction of higher dimensional packings is known only from numerical simulations. Currently, besides modelling adsorption monolayers, RSA packings are used as a starting point for molecular dynamics simulations of non-overlapping objects, as well as for protocols used in dense packings generations \cite{Lubachevsky1990,Torquato2010alg,Torquato2018hyperuniform}. Due to their properties, they define random meshes \cite{Ebeida2011delanuay,Ebeida2011voronoi}, which are used in computer graphics, to render high-quality images \cite{Pharr2016}.

It is worth to stress that RSA is also the simplest, but still a non-trivial model of random packing of hard objects~\cite{Zinchenco1994,Ghossein2013}, which accounts for excluded volume effects.  Therefore, there are many similarities between RSA packings and random close packings (RCP), where neighbouring particles have to touch themselves. For example, among the different packings built of ellipsoids, the densest one is observed for a very similar ratio of ellipsoid semi-axes length for both of these kinds of packings \cite{Donev2004,Man2005,Ciesla2019}.

Random sequential adsorption is also quite easy to implement, however, it has efficiency issues when packing becomes dense. In this case, the probability of finding randomly a large enough spot to place another molecule is very small, so a very large number of trials is needed. Moreover, it is never known if placing another particle is possible at all. Even if the number of successive failed attempts to add another object to a monolayer is huge, there is no guarantee, that the next trial will fail, too. These problems have already been solved for spherically symmetric molecules \cite{Wang1994,Ebeida2012,Zhang2013} and oriented rectangles \cite{Brosilow1991}. Recently, it has been shown how to effectively produce RSA packings built of rectangles \cite{Kasperek2018}, polygons \cite{Zhang2018}, ellipses, and spherocylinders \cite{Haiduk2018}. Although it allows to model quite many types of adsorption monolayers these shapes do not cover the whole possible spectrum of complex molecules. This paper presents how to effectively generate monolayers built of particles of shapes approximated by any number of spheres. These spheres may have various radii. Some of them may be disjoint, touching, or partially overlapped. To demonstrate how this method work, models of monolayers built of dimers \cite{Ciesla2014}, and fibrinogens \cite{Adamczyk2010} will be analysed.
\section{Algorithm}
\label{sec:algorithm}
\begin{figure}[ht]
  \centering{
    \includegraphics[width=0.6\columnwidth]{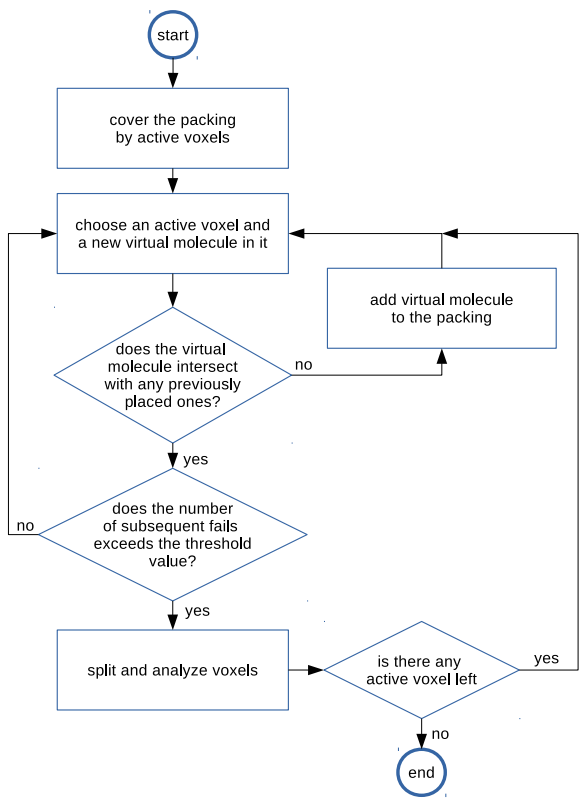}
  }
  \caption{Flowchart of the general algorithm to generate saturated random packing according to the random sequential adsorption protocol.} 
  \label{fig:flowchart}
\end{figure}

The algorithm is based on tracing regions where it is possible to place another molecule. A similar approach was applied previously for disks \cite{Wang1994, Ebeida2012, Zhang2013}, as well as anisotropic shapes \cite{Zhang2018, Haiduk2018, Kasperek2018}. At the beginning, the molecule can be placed anywhere, so the region for molecule placement is equal to the whole packing surface. Placing subsequent particles excludes part of it in this regard that there are no possibility to place there another molecule that will not intersect with any of previously placed particles. Therefore, subsequent virtual particles can be sampled only outside these excluded zones, which increases the probability of placing them. When the surface is fully covered by the excluded zones the algorithm stops as there is no possibility to place another molecule. The algorithm flowchart is shown in Fig.\ref{fig:flowchart}.
\subsection{Model of a complex molecule}
\begin{figure}[htb]
  \centering{
    \includegraphics[width=0.7\columnwidth]{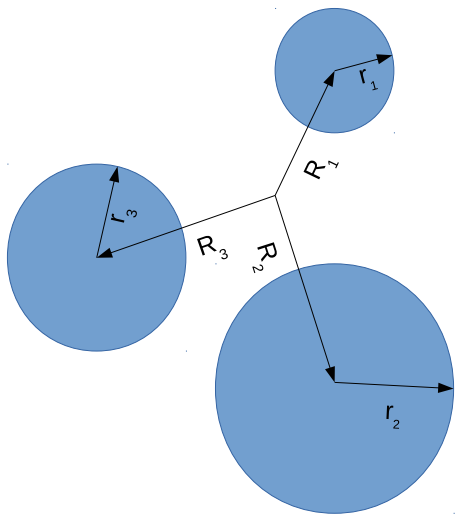}
  }
  \caption{Example of molecule approximated by three disks. The molecule shape is defined by vectors pointing to disks centres and its radii.} 
  \label{fig:example}
\end{figure}

It is assumed that the cross-section of a molecule and the surface can be approximated by $n$ disks of radii $\{r_1, r_2, ..., r_n\}$, with centres given by vectors $\{ {\bf R}_1, {\bf R}_2, ..., {\bf R}_n\}$. Example of a molecule is shown in Fig.\ref{fig:example}.
Positions and radii of the disks are arbitrary. They may be disjoint as in the figure, but they also may overlap. Origin of all the vectors ${\bf R}_i$ may also be at any arbitrary point.

\subsection{Voxels}
Each molecule that can be deposited on a surface is described by three coordinates: position of its centre and its orientation. Thus, the surface may be perceived as a three-dimensional object with two coordinates denoting particle position and the third being its orientation. At the beginning of the simulation, the surface is divided into disjoint voxels. Voxel $v$ is a cuboid in this three-dimensional space containing points $(x, y, \alpha ) \in [x_{v}, x_{v} + \Delta r) \otimes [y_{v}, y_{v} + \Delta r) \otimes [\alpha _{v}, \alpha _{v} + \Delta \alpha )$, where $(x_{v}, y_{v}, \alpha _{v})$ are voxel coordinates and $\Delta r$, and $\Delta \alpha $ are its dimensions. Voxel is inactive if there is no possibility to place a~new molecule in it because such a molecule will intersect with some of the previously placed particles. If a voxel becomes inactive it is removed because it no longer useful. To select random position and orientation of a~trial molecule, an active voxel and then a three-dimensional point inside it is chosen randomly.
\subsection{Intersection test} 
After selecting its position and orientation, the trial molecule is tested if it overlaps with any previously placed molecules on the surface. In this case, each disk of the trial molecule is checked if it intersects with any disk of other molecules. Two disks intersect when the distance between their centres is smaller than the sum of their radii. To speed up this test the cell method or the near-neighbour list method may be used, which allows checking disks of neighbouring molecules only \cite{Donev2005a, Donev2005b}.  
\subsection{Voxel division and elimination}
\label{sec:voxels}
When the number of subsequent unsuccessful trials to place a new particle on a surface reaches the given threshold value, all active voxels are divided. Each voxel of spatial size $\Delta r$ and angular size $\Delta \alpha $ is split into eight disjoint voxels of sizes $\Delta r/ 2$ and $\Delta \alpha /2$.

\begin{figure}
  \centering{
    \includegraphics[width=0.6\columnwidth]{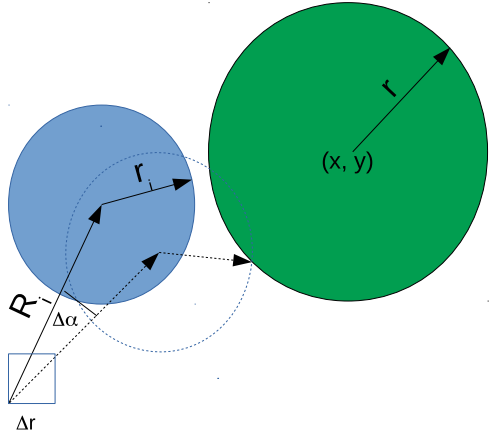}
  }
  \caption{Voxel intersection test. The green (darker) disk is a part of a molecule already added to the packing. The blue (lighter) disk is the $i$-th disk of a virtual particle, with its centre inside the voxel represented by the square. In this particular case, the voxel will not be eliminated because the maximal possible distance between the centres of the blue and the green disks is greater than $(r_{i} + r)$.}
  \label{fig:intersection}
\end{figure}

A voxel is removed when there is no possibility to place a new particle inside it, which means that for any position and orientation of the trial molecule inside the voxel, it will overlap with at least one disk belonging to molecule already placed on the surface. To find the condition describing such situation it is assumed that the voxel coordinates are $(x, y, \alpha )$. Thus, the position of the trial molecule origin inside this voxel is $(x + f_{x} \cdot \Delta r, y + f_{y} \cdot \Delta r, \alpha + f_{\alpha } \cdot \Delta \alpha )$, where $f_{x}, f_{y}$, and $f_{\alpha }$ are any numbers taken from $[0, 1)$ interval. For disk of radius $r$ with centre at $(x_{0}, y_{0})$, which belongs to a particle already placed on the surface -- see Fig.~\ref{fig:intersection}, the distance between this disk and the $i$-th disk of the trial molecule is:
%
\begin{equation}
d(f_{x}, f_{y}, f_{\alpha }) = d_{x}(f_{x}, f_{\alpha }) + d_{y}(f_{y}, f_{\alpha }),
\end{equation}
where
%
\begin{equation}
d_{x}(f_{x}, f_{\alpha }) = \left [ x + f_{x} \cdot \Delta r + R_{i} \cos \left (\alpha _{i} + f_{\alpha } \cdot \Delta \alpha \right ) - x_{0} \right ]^{2},
\end{equation}
and
%
\begin{equation}
d_{y}(f_{y}, f_{\alpha }) = \left [ y + f_{y} \cdot \Delta r + R_{i} \sin \left (\alpha _{i} + f_{\alpha } \cdot \Delta \alpha \right ) - y_{0} \right ]^{2}.
\end{equation}
$R_{i}$ and $(\alpha _{i} + f_{\alpha } \cdot \Delta \alpha)$ denote the length and the direction of the vector $\mathbf{R}_{i}$ pointing at the centre of $i$-th disk. The disks will intersect for all $(f_{x}, f_{y}, f_{\alpha})$ if the sum of maximal values of $d_{x}(f_{x}, f_{\alpha })$ and $d_{y}(f_{y}, f_{\alpha })$ is smaller than $(r_{i} + r)^{2}$. In such case, the voxel can be safely removed as it is not possible to place there a virtual molecule that does not intersect with the disk in $(x_{0},y_{0})$. To find the maximum of $d_{x}(f_{x}, f_{\alpha })$ it is necessary to determine the range of x-coordinate of the blue disk in Fig.~\ref{fig:intersection}, assuming that the beginning of vector $R_{i}$ is inside the voxel. It is given by
\begin{equation}
[x + R_{i} \cdot \min _{f_{\alpha } \in [0, 1)} \cos \left (\alpha _{i} + f_{\alpha } \cdot \Delta \alpha \right ),
\,\,\,
x + \Delta r + R_{i} \cdot \max _{f_{\alpha } \in [0, 1)} \cos \left (\alpha _{i} + f_{\alpha } \cdot \Delta \alpha \right )].
\end{equation}
The maximum of trigonometric function is either the maximum of $\cos \alpha $ and $\cos (\alpha + \Delta \alpha )$ or $1$, if $\alpha < 0 < \alpha + \Delta \alpha $ or $\alpha < 2\pi < \alpha + \Delta \alpha $. The maximum distance can be obtained by comparing $x_{0}$ with both ends of this interval. Similar analysis can be performed independently for $d_{y}(f_{y}, f_{\alpha })$.

It is possible to estimate maximum of $d(f_{x}, f_{y}, f_{\alpha })$ by adding maximal possible values of $d_{x}(f_{x}, f_{\alpha })$ and $d_{y}(f_{y}, f_{\alpha })$. If the sum is smaller than $(r_{i} + r)^{2}$ the voxel can be safely removed as it is not possible to place in it any new non-overlapping molecule.

The test is performed for all disks in the virtual molecule. Voxel is eliminated if any of the disks will intersect with any molecule that has been already added to the packing. Note that the voxel elimination test should be performed only for molecules, which are in the voxel's neighbourhood. Therefore, the same structures used to speed up molecules intersection test, like cell method or the near-neighbour method \cite{Donev2005a,Donev2005b} can be utilised here.

\section{Examples}
\begin{figure}[htb]
\vspace{0.3in}
  \centering{
    \includegraphics[width=0.7\columnwidth]{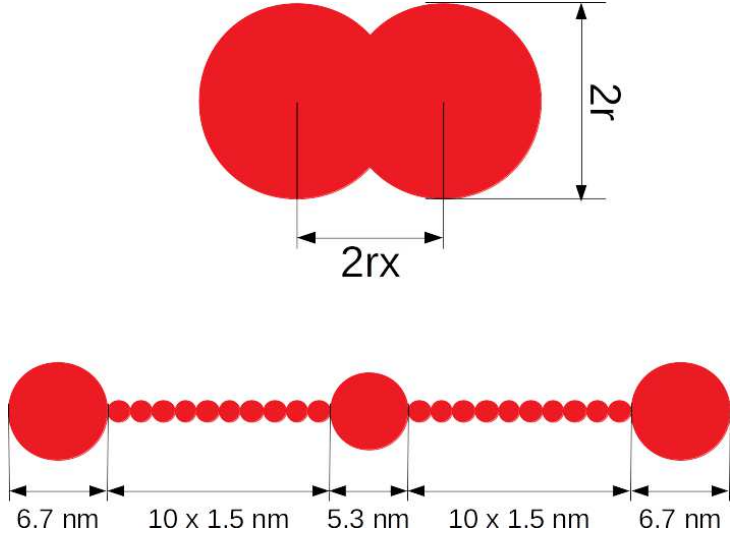}
  }
  \caption{Models of dimer (upper panel) and fibrinogen (lower panel) used to test the studied algorithm.} 
  \label{fig:dimer_fibrinogen}
\vspace{0.3in}
\end{figure}

The presented algorithm was used to determine saturated packing fraction for the model of dimer and fibrinogen particles (see Fig.\ref{fig:dimer_fibrinogen}). 
For each molecule, $100$ independent random monolayers were generated to determine the dependence of the mean packing fraction. The surface area was $S$ times bigger than the surface area of a single particle. To minimise the influence of finite size effects periodic boundary conditions were used \cite{Ciesla2018}. The mean packing fraction was calculated as $\theta = (1/100)\sum_{i=1}^{100} n_i/S$, where $n_i$ is the number of molecules in $i$-th monolayer. The statistical error of $\theta$ is given by $\sigma(\theta) = (1/100)\sqrt{\sum_{i=1}^{100} (n_i/S - \theta)^2}$.
\subsection{Dimers of different anisotropies}
\begin{figure}[htb]
\vspace{0.3in}
  \centering{
    \includegraphics[width=0.3\columnwidth]{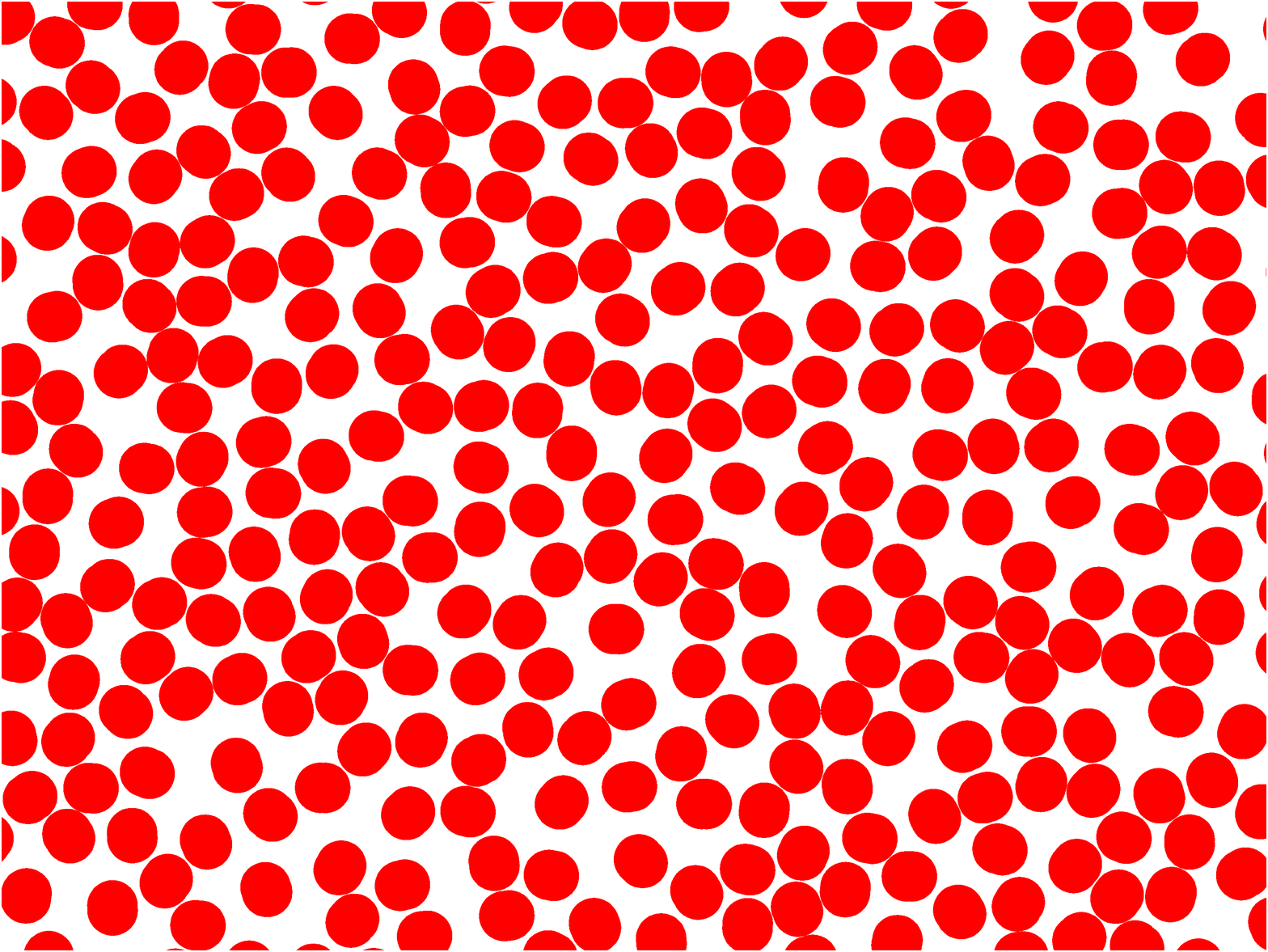}
    \includegraphics[width=0.3\columnwidth]{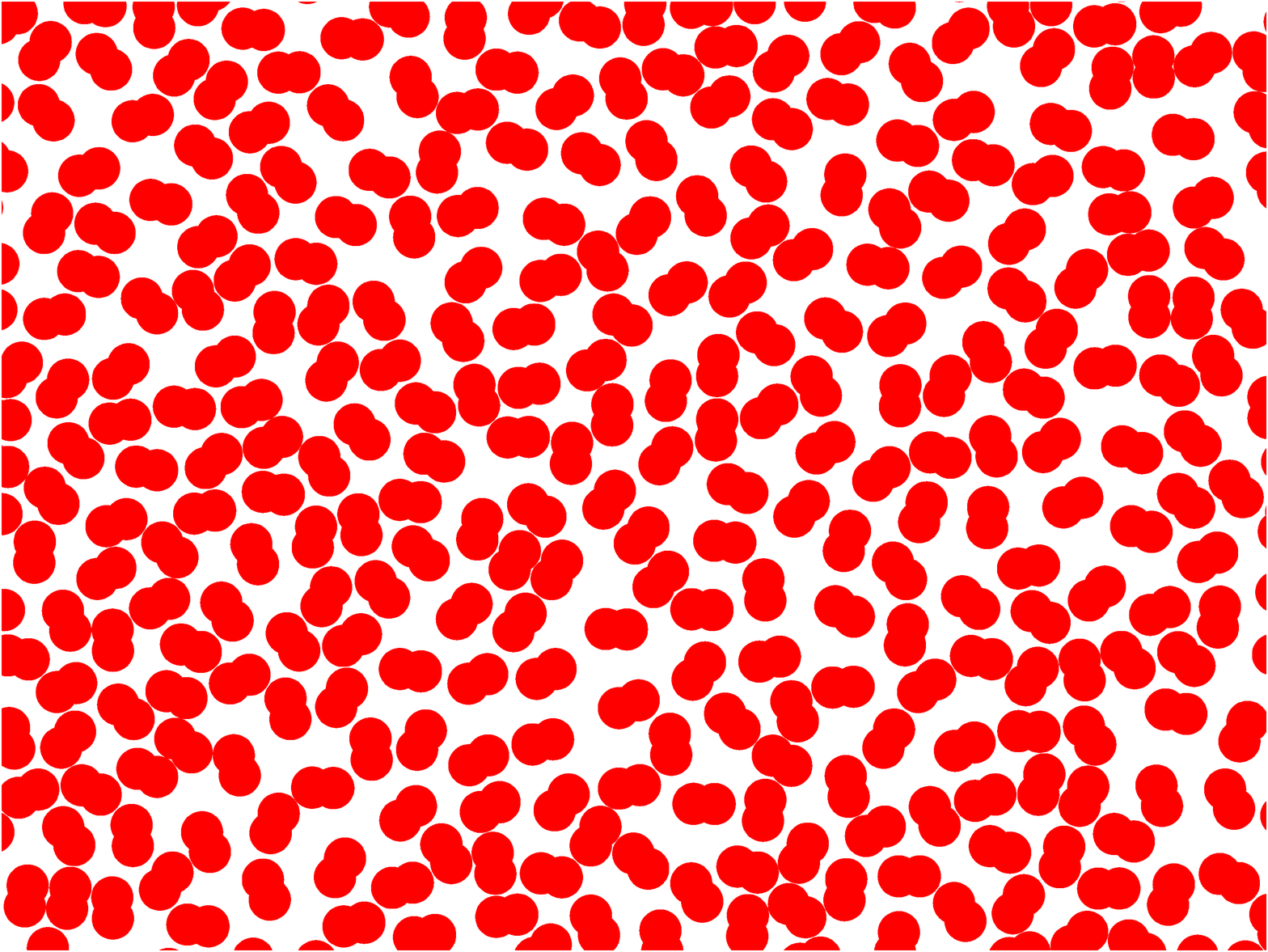}
    \includegraphics[width=0.3\columnwidth]{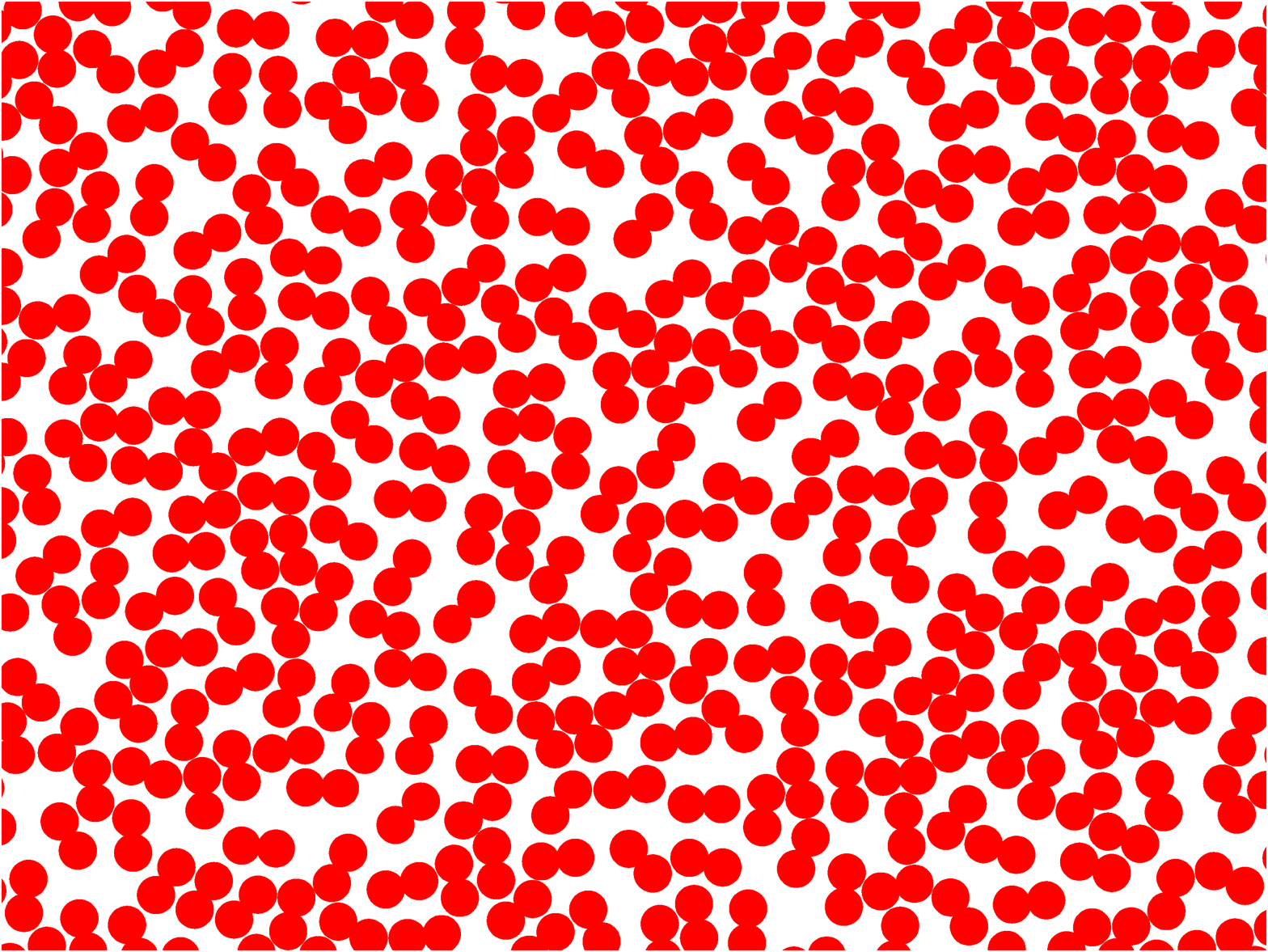}
  }
  \caption{Examples of saturated packing of dimers for $x=0.1$, $x=0.5$, and $x=0.9$ generated by the presented algorithm.} 
  \label{fig:dimers_example}
\vspace{0.3in}
\end{figure}

\begin{figure}[htb]
  \centering{
    \includegraphics[width=0.7\columnwidth]{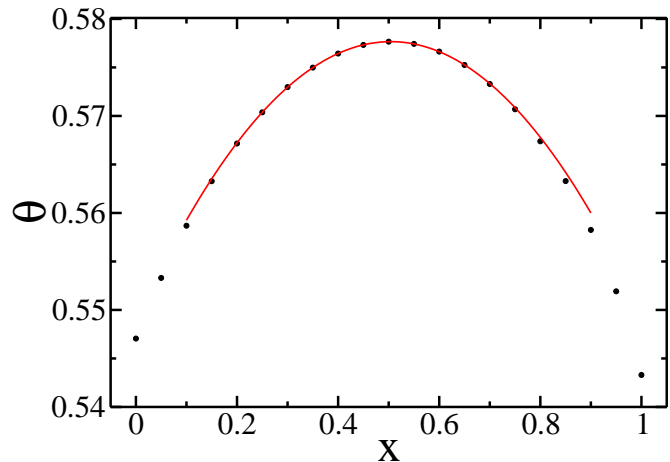}
  }
  \caption{Dependence of mean packing fraction of dimers on particle anisotropy $x$. Dots represent data obtained from numerical simulation, and the red solid line is a parabolic fit around the maximum: $\theta = -0.11273 \, x^2 + 0.11367 \, x + 0.54901$.} 
  \label{fig:dimers}
\end{figure}

Random sequential adsorption of such dimers may model various adsorption experiments \cite{Kujda2016, Salipante2016, Jachimska2018}, and gains increasing attention of theoretical studies \cite{Tarasevich2017, Budinski2017, Kundu2018}. In this section, the dependence of saturated packing fraction of dimers on their width-to-height ratio will be discussed \cite{Vigil1989, Ciesla2014, Ciesla2015}. The model of dimer is built of two identical, partially overlapped disks (see upper panel in Fig.\ref{fig:dimer_fibrinogen}). The width-to-height ratio is controlled by parameter $x$. In this study, it was assumed that $x \in [0, 1]$. For $x=0$ both disks fully overlap each other, and for $x=1$ they have only one common point. Fragments of obtained saturated random packings are shown in Fig.\ref{fig:dimers_example}.
The dependence of the mean packing density on parameter $x$, obtained for packing surface $S=10^6$ is shown in Fig. \ref{fig:dimers}.
As expected, the highest mean packing density is reached for moderate anisotropy \cite{Vigil1989, Ciesla2014}. Here, the optimal anisotropy corresponds to $x \approx 0.50$, for which the mean packing fraction there is $0.577658 \pm 0.000017$. The obtained result is similar to one reported previously \cite{Ciesla2014, Ciesla2015}, for studies where saturated packing fraction was estimated using kinetics of packing growth. In this study by using strictly saturated packing, the standard deviation of the mean packing density is twenty times smaller. Table \ref{tab:dimers} contains all the results from numerical simulations.  

\begin{table}[htb]
\label{tab:dimers}
\centering{
  \begin{tabular}{|c|c|c|}
  \hline
  x & mean packing fraction $\theta$ & standard deviation $\sigma(\theta)$\\
  \hline
  0.00 & 0.547053 & 0.000017 \\
  0.05 & 0.553301 & 0.000018 \\
  0.10 & 0.558686 & 0.000018 \\
  0.15 & 0.563276 & 0.000017 \\
  0.20 & 0.567160 & 0.000019 \\
  0.25 & 0.570372 & 0.000019 \\
  0.30 & 0.572973 & 0.000019 \\
  0.35 & 0.574987 & 0.000018 \\
  0.40 & 0.576433 & 0.000019 \\
  0.45 & 0.577321 & 0.000018 \\
  0.50 & 0.577658 & 0.000017 \\
  0.55 & 0.577428 & 0.000018 \\
  0.60 & 0.576637 & 0.000018 \\
  0.65 & 0.575254 & 0.000018 \\
  0.70 & 0.573296 & 0.000016 \\
  0.75 & 0.570686 & 0.000016 \\
  0.80 & 0.567386 & 0.000017 \\
  0.85 & 0.563285 & 0.000017 \\
  0.90 & 0.558254 & 0.000016 \\
  0.95 & 0.551921 & 0.000015 \\
  1.00 & 0.543301 & 0.000016 \\
  \hline
  \end{tabular}
}
\caption{Packing densities for saturated random packing of dimers for different particle anisotropy $x$.}
\end{table}
\subsection{Fibrinogen}
\begin{figure}[htb]
\vspace{0.3in}
  \centering{
    \includegraphics[width=0.7\columnwidth]{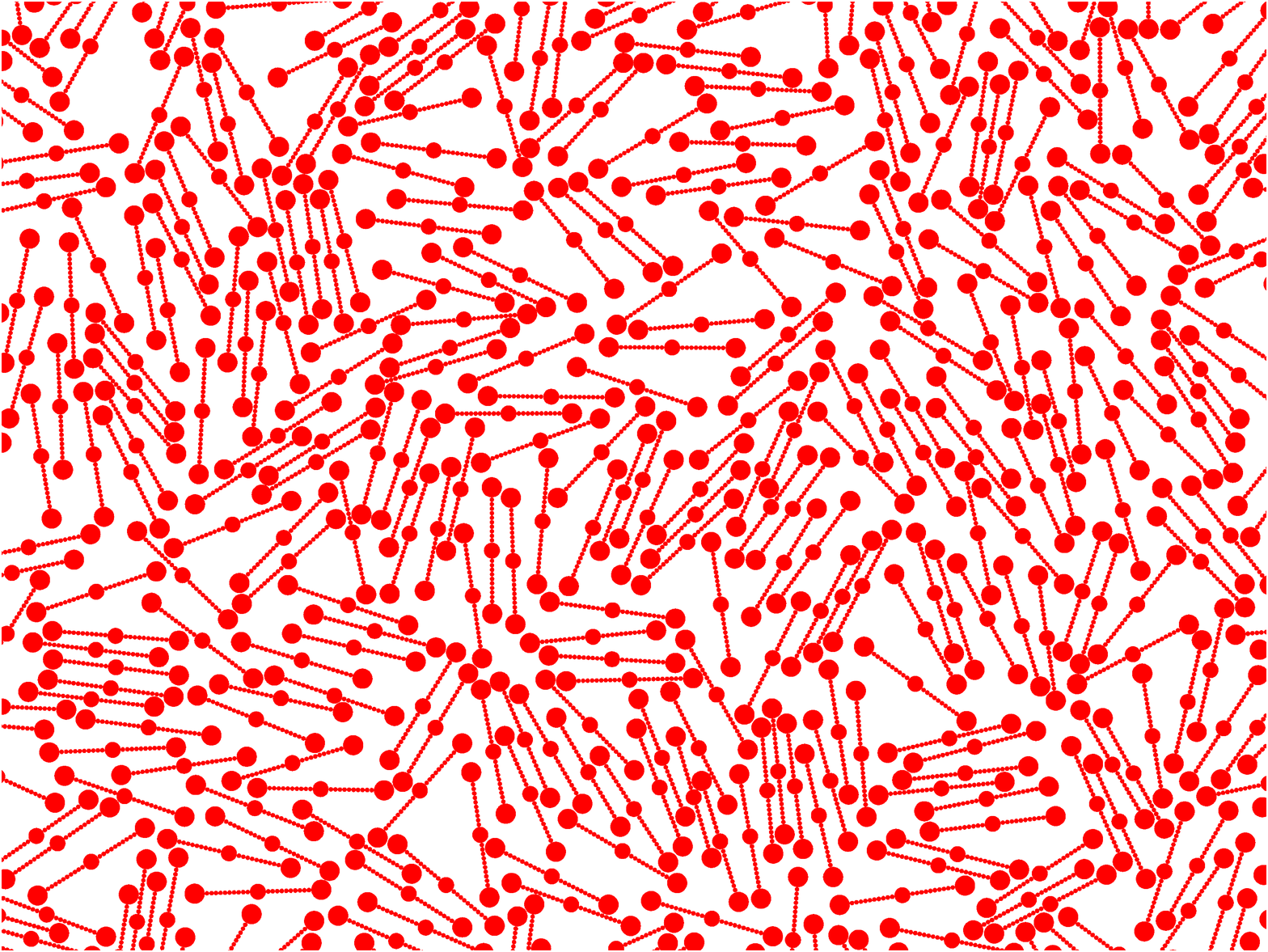}
  }
  \caption{Fragment of example saturated random packing of fibrinogen on a surface.} 
  \label{fig:fibrinogen}
\vspace{0.3in}
\end{figure}

The model of fibrinogen particle \cite{Adamczyk2010} consists of 23 disks (see lower panel in Fig.\ref{fig:dimer_fibrinogen}). Example of saturated random packing for such molecules is shown in Fig.\ref{fig:fibrinogen}.
In this case, due to much complex molecule shape, the packing size was generated in the simulation was hundred times smaller than for dimers and equal to $S=10^4$. The calculated value of the mean saturated packing fraction was $\theta = 0.29633 \pm 0.00018$. Again, the obtained value agrees with the one reported earlier \cite{Adamczyk2010}, but is much more accurate.
\section{Summary}
The study presents an algorithm, which allows to generate saturated random sequential adsorption packings of molecules built of disks. The algorithm can be used for modelling monolayers built during irreversible adsorption of complex molecules. As it traces regions where subsequent particles may be placed, it is much faster than classical RSA method and it stops when there is no possibility to add another particle to the packing.
\section*{Acknowledgements}
This work was supported by grant no. 2016/23/B/ST3/01145 of the National Science Center, Poland. Numerical simulations were carried out with the support of the Interdisciplinary Center for Mathematical and Computational Modeling (ICM) at University of Warsaw under grant No.\ GB-76-1.


\bibliographystyle{model1-num-names}
\bibliography{main.bib}   

\end{document}